\documentstyle[twocolumn,floats,graphicx,aps,prl]{revtex}
\begin{document}
%\preprint{JFI-xx-xx-xx}
%\draft %command makes pacs numbers print
\wideabs{
\title{Determining Pair Interactions from Structural Correlations}
\author{Muhittin Mungan, Chorng-Haur Sow, Susan N. Coppersmith, 
and David G. Grier}
\address{The James Franck Institute and Department of Physics\\
The University of Chicago, 5640 S. Ellis Ave., Chicago, IL 60637}
\date{\today}
\maketitle
\begin{abstract}
 We examine metastable configurations of a
 two-dimensional system of interacting particles
 on a quenched random potential landscape and ask how the configurational
 pair correlation function is related to the particle interactions
 and the statistical properties of the potential landscape. 
 Understanding this relation facilitates  
 quantitative studies of magnetic flux line interactions in  
 type II superconductors, using structural information available 
 from Lorentz microscope images or Bitter decorations.
 Previous work by some of us supported the conjecture that
 the relationship between pair correlations and interactions in
 pinned flux line ensembles is analogous to the corresponding
 relationship in the theory of simple liquids.
 The present paper
 aims at a more thorough understanding of this relation. We report
 the results of numerical simulations and present a theory for the
 low density behavior of the pair correlation function which agrees 
 well with our simulations and captures features   
 observed in experiments. In particular, we find
 that the resulting description goes beyond the conjectured classical liquid
 type relation and we remark on the differences.         
\end{abstract}
} % end of wideabs

\section{Introduction}
Magnetic flux lines in type II superconductors belong to a class of systems 
in which interacting particles are subjected to 
random pinning forces \cite{G1,G2}. 
At very low temperatures, the resulting
configurations are determined by a competition between elastic energy cost
and pinning energy gain as the flux line ensemble attempts 
to accommodate the random medium. 
This competition gives rise to a rich and complex  energy landscape 
of metastable states. 
One expects that structural properties of the particle configurations 
should contain information about the
interparticle interactions as well as the statistical properties of
the pinning potential landscape, the {\em pinscape}. 

\begin{figure}[htbp]
  \begin{center}    
    \includegraphics[width=3in]{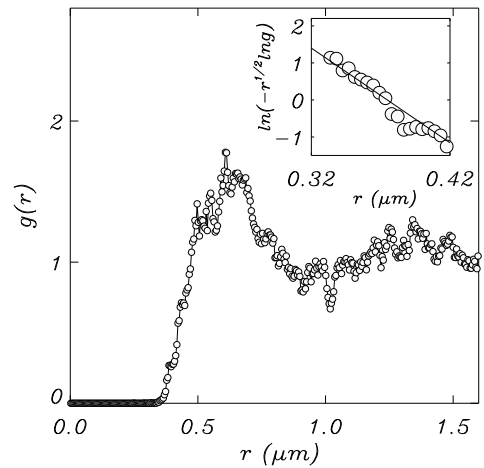}
        \caption{Pair correlation function obtained
        from averaging over Lorentz microscope images of
        flux lines in Nb \protect\cite{SHTCG,Harada} at $T = 5$ K.
        The region of interest is where $g(r)$ rises from zero
        ({\em lift-off}). The inset shows a comparison of the
        lift-off region of $g(r)$ with liquid theory behavior,
        Eq.~(\protect\ref{eqn:liq}) (refer to section II for further
        details).}
    \label{fig:fig1}
  \end{center} 
\end{figure}

In an earlier publication \cite{SHTCG} by some of us, henceforth
referred to as SHTCG, the near-equilibrium configurations
of flux lines in a Nb thin film obtained from Lorentz microscope images
\cite{Harada} were analyzed. SHTCG argued that a sequence of 
flux line configurations on the fixed pinscape resembles 
the instantaneous configurations
of a classical simple liquid, with the underlying pinscape playing the role
of an effective heat bath. 
This conjecture implies that in the low-density limit, the pair-correlation
function $g(r)$ should be of the form \cite{H&MD}
\begin{equation}
g(r) = e^{-U(r)/U_o} + \ldots,
\label{eqn:liq}
\end{equation}
where $U(r)$ is the flux line pair interaction, $U_o$ 
is an effective pinning energy scale,
and the terms omitted correspond to higher order corrections due to
many-body effects. 
For stiff magnetic flux lines, the interparticle potential $U(r)$ 
(per unit length) is given in the
low density limit by the asymptotic form 
\cite{deGennes,Abra}
\begin{equation}
U(r) = \frac{{\phi_{o}}^2}{8 \pi^2 \lambda^2} \sqrt{\frac{\pi \lambda }{2r}} 
\, e^{-r / \lambda},
\label{eqn:Kasympt}
\end{equation} 
where
$\phi_{o}$ is the flux quantum 
and $\lambda$ is the London penetration depth. 
By treating $U_o$ as an adjustable parameter,
SHTCG showed that the small $r$ behavior ({\em lift-off})
of the pair correlation function 
agrees well with the form given in Eq.~(\ref{eqn:liq}),
%. They thus verified the functional form of the pair interactions 
yielding a value
for the penetration depth in good agreement with the accepted value for Nb
(see Fig.~\ref{fig:fig1}). 
Bitter decoration images of 
flux line configurations in the high temperature superconductor 
BSCCO analyzed in the same manner 
also yield results in good quantitative agreement with the accepted 
value of the penetration depth for BSCCO \cite{us}. The fit value for
$U_o$ presumably describes the characteristic energy scale for
the pinscape disorder.

        The present paper aims at a more thorough understanding of 
how structural correlations arise from microscopic properties
such as the particle interactions and potential landscape.
Particular questions we address include: 
Under what conditions can the static configurations of an 
interacting system in a quenched random potential give rise to liquid-like
correlations of the form Eq.~(\ref{eqn:liq})? How does Eq.~(\ref{eqn:liq}) 
compare 
to a theory where the quenched 
nature of randomness is accounted for explicitly?

We report the results of numerical simulations 
and present a theory that captures the essential
features observed both in experiments and in our simulations, providing thus
some answers to the questions raised above.
In particular our simulations show that
a relation of the form Eq.~(\ref{eqn:liq}) emerges in the limit of point-like
pins, {\em i.e.} pins of small spatial extent, 
and for broad ({\em e.g.} exponential) pinning strength distributions.
In contrast, pinscapes composed 
of identically strong or spatially extended pins 
are not well described by Eq.~(\ref{eqn:liq}). Our theory
is based on the observation that the 
behavior of the pair correlation function at small $r$ is
dominated by strongly pinned flux lines. In the low-density limit,
the functional form of the
pair correlation function therefore arises essentially 
from a convolution of the tails
of the pinning energy distribution with the probability distribution for 
locating a flux line at a given position. 

The paper is organized as follows. We present our model and
its motivation in Section II. Section III contains details of the
numerical simulations. Numerical results and their
discussion are given in Section IV, while Section V is devoted to a 
description of the underlying theoretical
picture. We conclude with a discussion and summary of our results in 
Section VI.

\section{The Model}

        We consider a two-dimensional system of $N$ particles 
in a quenched random 
pinscape consisting of discrete pinning sites positioned at random.
The particles, which can be interpreted as
stiff magnetic flux lines, have positions that are 
specified by (two-dimensional) vectors ${\bf r}_i$. 
As a convention, particle and pinning site locations 
are subscripted with Latin and Greek indices, respectively.
The force on the $i^{th}$ flux line at position ${\bf r}_i$ is given by
\begin{equation}
        {\bf F}_i = - \sum_{j \neq i}^{N} {\bf \nabla}_i 
                        U({\bf r}_i - {\bf r}_j)
                        - \sum_{\alpha} {\bf \nabla}_i V_{\alpha}({\bf r}_i - 
                                {\bf r}_{\alpha}).
\label{eqn:Fsubi}
\end{equation}
Here the first and second sum are the contributions to the 
total force on a given particle due to its
interactions with the other particles and with the pinning sites;
$U(r)$ is the interparticle interaction, whereas $V_\alpha(r)$
is the particle-
pin interaction. 
For stiff magnetic flux lines, the interparticle potential $U(r)$ 
(per unit length) is given in the
low-density limit by the asymptotic form Eq.~(\ref{eqn:Kasympt}).
We assume that the pinning potential due to a single pin is
attractive and short ranged, 
\begin{equation}
V_{\alpha}(r) = - V_{\alpha} e^{-r^2/r_p^2},
\label{eqn:pinv}
\end{equation}
where $V_{\alpha}$ is the pin's strength, 
and $r_p$ its range.

The locations of the pinning sites ${\bf r}_\alpha$ 
are assumed to be distributed 
at random. We seek stable and static flux line configurations 
$\{ {\bf r}^{*}_i \}$ given by solutions to
\begin{equation}
        {\bf F}_i({\bf r}^*_1,{\bf r}^*_2, \ldots,{\bf r}^*_N) = {\bf 0},
\label{eqn:Feqz}
\end{equation}
with the additional stability constraint that the matrix 
$-\partial{{\bf F}_i}/\partial{{\bf r}^*_j}$ be positive-definite.  
If $\{ {\bf r}^*_1,{\bf r}^*_2, \ldots,{\bf r}^*_N \}$ is such a 
configuration, the pair correlation function
$g^*(r)$ for this configuration is given by
\begin{equation}
        g^*(r) = \frac{\langle \int \rho^* ({\bf x}-{\bf r})
          \rho^*({\bf x}) \, d{\bf x} \rangle_\theta}
                      {[\int \rho^* ({\bf x}) \, d{\bf x}]^2},
\label{eqn:gstar}
\end{equation}
with
\begin{equation}
  \rho^*({\bf x}) = \sum_{i=1}^N \delta({\bf x} - {\bf r}^*_i),
%\nonumber
\end{equation}
and $\langle \ldots \rangle_\theta $ denoting an average over angles.
The pair correlation function $g(r)$ is obtained by averaging 
Eq.~(\ref{eqn:gstar}) over independent configurations.

If the analogy to liquid structure theory adequately describes the
emergence of structure in this quenched random system, then
Eqs.~(\ref{eqn:liq}) and (\ref{eqn:Kasympt}) imply that (to lowest order)
\begin{equation}
        \ln (-r^{\frac{1}{2}} \ln g(r)) \propto - \frac{r}{\lambda}.
\label{eqn:lnliq}
\end{equation}
Thus liquid theory predicts that 
a plot of the corresponding quantity versus $r$ should yield a linear
relation  with slope $1/\lambda$, as observed experimentally
in the inset to Fig.~\ref{fig:fig1}.

Our numerical work involves 
solving Eq.~(\ref{eqn:Feqz}) 
for static configurations and thereby obtaining the configuration-averaged
pair correlation function. Details of the implementation 
and presentation of our numerical results are given in the following
two sections.

\section{Numerical Implementation}
In this section we describe our numerical implementation. 
Simulations were carried out on a 
square region with $7250$ randomly placed
pins.   
Having constructed a random pinscape,  
we placed $N = 65$ flux lines at random and 
solved numerically for stable equilibrium configurations of
Eq.~(\ref{eqn:Feqz})
using a three stage procedure: (i) Starting from a random initial
configuration, the vortex system is allowed to relax dynamically 
according to
\begin{equation}
        \frac{d{\bf{r}}_i}{dt} = {\bf F}_i({\bf r}_1,{\bf r}_2, \ldots,
                                {\bf r}_N),
\end{equation}
until the magnitudes of all forces on the RHS are less than a prescribed
target value.
(ii) Once this has been achieved, the resulting 
configuration is
used as an initial guess in a Newton-Raphson algorithm to find 
a solution to Eq.~(\ref{eqn:Feqz}).
Solutions
found this way are not guaranteed to be stable with respect
to small perturbations of the particle locations. Therefore, (iii),
a linear stability analysis is carried out to check for stability. If the
configuration obtained this way does not turn out to be stable, stage (i)
is resumed with a reduced target value.
The whole procedure is repeated until a stable 
configuration has been found.
The
combination of a molecular dynamics
algorithm with a Newton-Raphson scheme, as described above, turns out to be
far more computationally efficient than using molecular 
dynamics alone \cite{AAM}. 

Lengths were chosen so that the square corresponds to a 
$5 \times 5~\mu$m$^2$ field of view in a magnetic field of  roughly $50$ Gauss.
With this choice, all lengths will be reported in $\mu$m.
The London penetration depth was taken to be $\lambda = 0.04~{\mu}$m.
These choices resemble the experimental condition in Ref.~\cite{SHTCG}.  
The parameters varied in the simulations were the 
range of the pinning forces $r_p$, with values $r_p \in [0.015,0.05]~\mu$m,
and the distribution of pinning
strengths $V_\alpha$, {\em cf.} Eq.~(\ref{eqn:pinv}).
We considered the following distributions: 
\begin{equation}
        {\cal P}(V_\alpha) = \delta(V_\alpha - V_o),
\label{eqn:dpin}
\end{equation}
for {\em identical} pinning strengths, and
\begin{equation}
        {\cal P}_\nu (V_\alpha) = \frac{c_1}{V_o} 
                e^{-(c_2 \frac{V_\alpha}{V_o})^\nu}, V_\alpha > 0,
\label{eqn:expin}
\end{equation}
for stretched-exponentially distributed pinning
strengths, with the constants $c_1$ and $c_2$ given by 
$c_1 = \nu c_2/\Gamma(1/\nu)$,  and $c_2 = 
\frac{\Gamma(2/\nu)}{\Gamma(1/\nu)}$, 
where $\Gamma(x)$ is the Gamma function. 
In particular, the distributions used were
$\nu = 2$ ({\em half-gaussian}),
$\nu = 1$ ({\em exponential}),  and $\nu = 1/2$ 
({\em stretched exponential}).
The mean of the above distributions is given by $V_o$.
The unit of energy was chosen 
such that
\begin{equation} 
 \frac{{\phi_{o}}^2 r_p}{8 \pi^2 \lambda^2 V_o} = 200.
\end{equation}
This choice is consistent with experimental observations
\cite{SHTCG}, and ensures that the maximum force exerted
by a pin is independent of the range of the pinning potential.

The numerical integration in stage (i) was carried out using a $4^{th}$
order Runge-Kutta method \cite{NumRec} with adaptive stepsize 
and an initial force resolution of $10^{-6}$ in the above units
with subsequent decrements by factors of 10, as needed.
Algorithms of the software libraries
LINPACK and EISPACK \cite{Pack} were used in putting 
together stages (ii) and (iii).
The maximum residual force on any particle in a stable configuration
was found to be smaller than $\Delta F = {\cal O}(10^{-12})$.

Free boundary conditions were employed, and flux lines leaving the 
square (leaks) were
put back in random locations. Typically, only a few 
leaks occurred in a given run.

In order to speed up computations further, several additional
measures were taken: If a stable configuration 
was not found after a fixed number of 
integration steps $(5000)$, 
a new initial configuration was generated. Initial
configurations were such that flux lines were randomly positioned, but 
no two were allowed to be closer initially than $0.2~\mu$m. 
These measures were found to 
reduce the computation time without 
changing the results significantly. 
The particle-pin interaction, Eq.~(\ref{eqn:pinv}),
was cut off at a distance of $4r_p$. Prior to each run, the  
area was subdivided into a mesh of
squares and a look-up table was generated providing each square with the
location and 
strength of pins within the effective interaction distance $4r_p$. This
eliminates the need to search all pinning sites for each computation of 
the pinning forces. 
\begin{figure}[tbhp]
  \begin{center}    
    \includegraphics[width=3in]{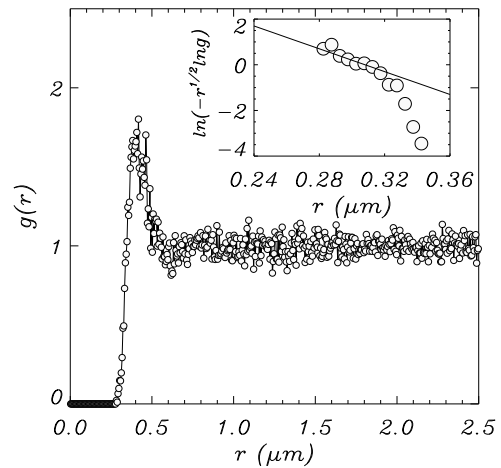}
    \caption{Pair correlation function $g(r)$ 
        obtained from averaging over $212$ stable
        flux line configurations on a pinscape, 
        for $\lambda = 0.04$ and $r_p = 0.015$.
        The pinning strengths are exponentially
        distributed. The inset shows a comparison with the 
        liquid theory prediction, Eq.~(\ref{eqn:lnliq}).}
    \label{fig:fig2}
  \end{center}
\end{figure}

\section{Numerical Results}
In this section we present the results of our simulations described above.
Fig.~\ref{fig:fig2} shows the scaling  of the pair correlation function
obtained from averaging over $212$ stable flux line configurations on
a pinscape generated with exponentially distributed pinning strengths 
according to 
Eq.~(\ref{eqn:expin}), with $r_p = 0.015$. 
The inset to Fig.~\ref{fig:fig2} 
shows the scaling behavior of $ \ln (-r^{\frac{1}{2}} \ln g(r)) $ with
$r$. The solid line has slope $1/\lambda$. Note the good agreement with
the liquid theory prediction, Eq.~(\ref{eqn:lnliq}), for values of $r$ in the
initial lift-off of $g(r)$. The deviations for larger $r$ arise from
excess correlations ({\em pile-up}) near the peak of $g(r)$ not 
accounted for in the na\"{\i}ve scaling Ansatz. Such excess correlations 
also are seen experimentally \cite{SHTCG,us}. 

\begin{figure}[htbp]
  \begin{center}    
    \includegraphics[width=3in]{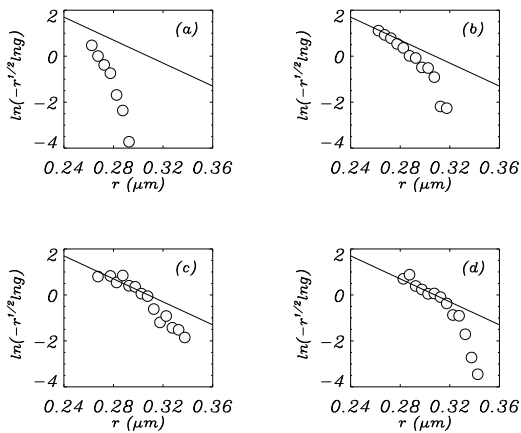}
    \caption{Dependence of the pair correlation 
        function on the range of the 
        pinning potential $r_p$ for exponentially 
        distributed pinning strengths. 
        Solid lines correspond to a liquid theory 
        scaling, Eq.~ (\ref{eqn:lnliq}), 
        with $\lambda = 0.04~\mu$m. The values for $r_p$ 
        are $0.05~\mu$m (a), $0.025~\mu$m (b), $0.017~\mu$m (c), 
        and $0.015~\mu$m (d). Averages were calculated 
        from $596$, $600$, $323$ and $212$ configurations, 
        respectively.}
    \label{fig:fig3}
  \end{center}
\end{figure}

Having thus observed numerically a liquid-like relation of the form 
Eq.~(\ref{eqn:lnliq}),   
we next ask how the behavior of the 
pair correlation function depends on
the range of the pinning potential 
$r_p$. 
Fig.~\ref{fig:fig3} shows the scaling 
of $ \ln (-r^{\frac{1}{2}} \ln g(r)) $ with
$r$ for different values of $r_p$,  ranging 
from $0.015~\mu$m to $0.05~\mu$m. 
The asymptotic form, Eq.~(\ref{eqn:lnliq}), 
is realized only in the limit of a
short-ranged pinning potential ($r_p \ll \lambda$). 

\begin{figure}[htbp]
  \begin{center}    
    \includegraphics[width=3in]{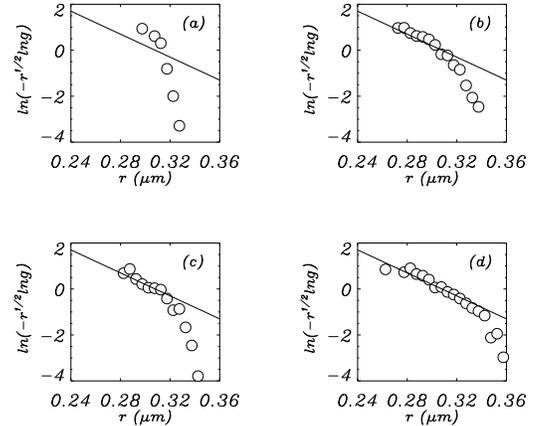}
    \caption{Dependence of the pair correlation 
        function on the distribution
        of pinning strengths: (a) identical pinning strengths, 
        (b) half-gaussian, $ \nu = 2$, (c) exponential, 
        $\nu = 1$, and (d)
        stretched exponential, $\nu = 1/2$. Solid
        lines indicate the scaling, Eq.~(\ref{eqn:lnliq}), 
        for $\lambda = 0.04$ ($r_p = 0.015$). Averages 
        were obtained from $148$, $228$, $212$, and $152$
        configurations, respectively.}
    \label{fig:fig4}
  \end{center}
\end{figure}

The dependence of the pair correlation 
function on the breadth of the pinning 
strength distribution is shown in Fig.~\ref{fig:fig4}. Here we compare the
results for configurations on a pinscape 
of spatially narrow pins whose strengths 
were drawn from increasingly broad 
distributions: delta function 
[Eq.~(\ref{eqn:dpin})], $\nu = 2$, $1$ and $1/2$ [Eq.~(\ref{eqn:expin})]. 
The pair correlation functions corresponding to the three broad 
distributions, $\nu = 2$, $1$ and $1/2$ agree with the prediction of
the liquid-like theory, Eq.~(\ref{eqn:liq}), at least for small $r$.
In contrast, Eq.~(\ref{eqn:liq}) does not describe the case of identical 
pinning
strengths; the rapid fall-off of $ \ln (-r^{\frac{1}{2}} \ln g(r)) $
indicates a steeper than expected lift-off of $g(r)$.

        Thus the results of our numerical simulations clearly show a 
liquid-like behavior
of the form Eq.~(\ref{eqn:lnliq}) as observed in experiments. Furthermore
our simulations indicate that this scaling only
arises in the limit of point-like pins drawn
from a broad distribution of pinning strengths. Its successful application
to experimental data suggests that the pinscape in these materials
also is characterized by point-like pins with a broad distribution
of strengths. Moreover, these observations serve as a warning that
simulation results based on identical 
pinning strengths may differ subtly from the behavior of physical
systems.

\section{Theoretical Description}
The analogy drawn between pinned flux lines and thermally disordered
fluids yields an empirically satisfactory description of flux line
pair correlations. To this point, though, it has been justified by 
heuristic arguments. This section introduces a statistical treatment
of structure in strongly pinned flux lines which provides
a superior description of disordered flux line distributions yet yields
the liquid structure result, Eq.~(\ref{eqn:liq}), as a limiting case.

We begin by assuming that the dominant contribution to the force on a given
flux line is due to its nearest neighbor.
Because of the short-range nature
of the flux line interactions, this assumption is justified in the limit of
low flux line concentration, where the average flux line separation, as
well as fluctuations around it, are much larger 
than the London penetration depth. Furthermore, our simulations 
reveal that more than $99 \%$ of the flux lines in a stable configuration 
are {\em pinned}, {\em i.e.} within a distance
of $r_p/\sqrt{2}$ of a pinning site. Thus assuming that the flux lines
are pinned in pairs is reasonable. 
In particular, this approximation is valid 
for pairs of flux lines pinned at distances smaller than average.
These pairs
determine the behavior of the pair correlation function near lift-off, 
{\em i.e.\ }the region where the function rises from zero.
They occur only if 
sufficiently strong pins are available.  In other words, a pair of 
nearest-neighbor flux lines can be 
pinned at a separation $r$ only if both
pins are strong enough to overcome the flux line repulsion force, $f(r) = 
-\partial U(r)/\partial r$. 

From Eq.~(\ref{eqn:pinv}) we see that a pin $\alpha$ of strength
$V_\alpha$ can exert a maximum force given by
\begin{equation}
        F_\alpha = \sqrt{\frac{2}{e}} \: \frac{V_\alpha}{r_p}.
\label{eqn:fmax}
\end{equation}
Let $P_d(F_\alpha)$ be the probability that a flux line has come to rest
on a pin of maximal pinning force $F_\alpha$.
The cumulative probability
\begin{equation}
        Q_d(F) = \int_{F}^{\infty} P_d(F_\alpha) dF_\alpha.
\end{equation}
is the likelihood that the occupied pin is at least as strong as $F$.
 If we assume that flux lines are
equally likely to be pinned anywhere in the sample 
as long as the corresponding pinning forces
are large enough to overcome the flux line interaction, 
it follows that the
pair correlation function $g(r)$ has the form
\begin{equation}
        g(r) = {Q_d}^2(f(r)) + \ldots .
\label{eqn:gth1}
\end{equation}
        This expression should be valid in the low-density
limit, where only pairwise-pinned flux lines have to be considered 
and screening effects due to the presence of other flux line
pairs can be neglected. One can readily show that 
the resulting flux line configurations are stable.
 
\begin{figure}[htbp]
  \begin{center}    
    \includegraphics[width=3in]{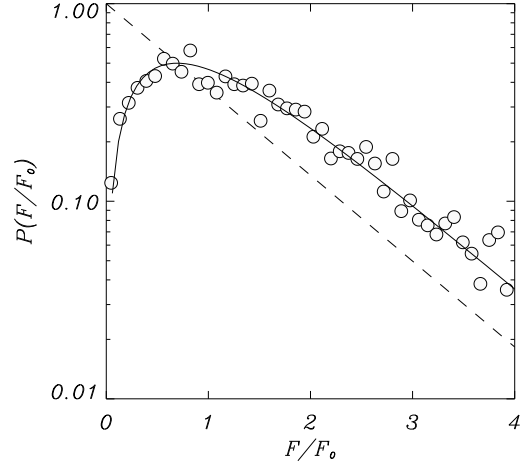}
    \caption{Distribution of available and 
        dynamically selected maximal 
        pinning forces. Shown here is the case of 
        exponentially distributed pinning
        strengths, $\nu = 1$. The dashed line 
        corresponds to the distribution of 
        available maximal pinning forces as obtained from 
        Eqs.~(\ref{eqn:fmax}) and
        (\ref{eqn:expin}). Data points correspond 
        to the distribution obtained by
        determining the pinning strength of the pins 
        associated with every pinned 
        flux line. The solid line corresponds to 
        Eq.~(\ref{eqn:pdexp}), 
        as explained in the text.}
    \label{fig:fig5}
  \end{center}
\end{figure}

The distribution of dynamically 
selected pinning strengths, $P_d(F_\alpha)$, does not 
coincide with the corresponding 
distribution of {\em a priori} available 
strengths, $P(F_\alpha)$. The latter is
readily obtained from
the pinning energy distribution, ${\cal P}_\nu(V_\alpha)$, 
using Eq.~(\ref{eqn:fmax}).
Fig.~\ref{fig:fig5} compares the distributions 
of dynamically
selected and available maximal pinning forces for exponentially distributed 
($\nu = 1$) pinning strengths, Eq.~(\ref{eqn:expin}). 
We see clearly that the particles preferentially occupy
strong pins. This is consistent with our expectation that
an isolated flux line, 
or one that is effectively isolated, is more likely
to be in the domain of attraction of a strong pin in its vicinity. 

On the basis of this observation, we hypothesize that a given flux line
dynamically selects the strongest of $n$ pins in its immediate
vicinity so that
\begin{equation}
        Q_d(F) = 1 - \left [1 - Q(F) \right ]^n,
\label{eqn:qd}
\end{equation}
and thus
\begin{equation}
        P_d(F) = n P(F) \left [\int_{0}^{F} P(F')dF'\right ]^{n-1}.
\label{eqn:pd}
\end{equation} 
Equations~(\ref{eqn:gth1}) -- (\ref{eqn:pd})
and their successful application to our simulations
are the central results of this article.

Fig.~\ref{fig:fig5} 
shows that $P_d(F)$ is very well approximated by $n = 2$, {\em i.e.},
\begin{equation}
        P_d(F) = 2 P(F) \int_{0}^{F} P(F')dF',
\label{eqn:pds}
\end{equation} 
implying that its cumulative distribution $Q_d(F)$ obeys, 
\begin{equation}
        Q_d(F) = 1 - \left [1 - Q(F) \right ]^2.
\label{eqn:qds}
\end{equation}
We find that $n=2$ also satisfactorily describes
dynamically selected pinning in our simulations with half-gaussian 
and stretched exponentially distributed 
pinning strengths, $\nu = 2$ and $\nu = 1/2$.

For exponentially distributed pinning 
strengths, $\nu = 1$, we obtain
\begin{equation}
        P_d(F) = \frac{2}{F_o} e^{-F/F_o}
                \left (1 - e^{-F/F_o} \right ),
\label{eqn:pdexp}
\end{equation} 
where $F_o = \sqrt{2/e} \: V_o / r_p$.
Thus,
\begin{equation}
        Q_d(F) = 2 e^{-F/F_o}\left (1 - \frac{1}{2} e^{-F/F_o} \right ),
\label{eqn:qdexp}
\end{equation}
and, using Eq.~(\ref{eqn:gth1}),
\begin{equation}
        g(r) = 4 e^{-2f(r)/F_o}
                \left [1 - \frac{1}{2} e^{-f(r)/F_o}\right ]^2 + \ldots.
\label{eqn:gthexp}
\end{equation} 
Analogous predictions can be obtained for the cases 
$\nu = 2$ and $\nu = 1/2$. 

\begin{figure}[htbp]
  \begin{center}    
    \includegraphics[width=3in]{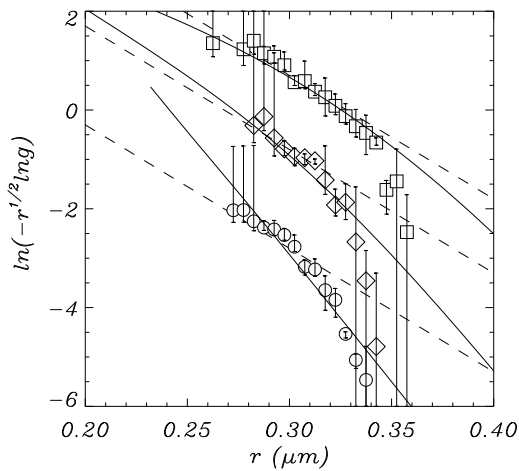}
    \caption{Comparison of the behavior of the pair 
        correlation function near lift-off against 
        simulation results, liquid theory type scaling
        (dashed lines) and theoretical prediction, 
        Eqs.~(\protect\ref{eqn:gth1}) and (\protect\ref{eqn:qds}), for 
        half-gaussian (circles), exponential (diamonds), 
        and stretched exponential $\nu = 1/2$ (boxes) 
        distribution of pinning strengths. Note that the 
        theoretical predictions contain no free
        parameter once $n$ has been fixed (Fig.~\protect\ref{fig:fig5}), 
        while the liquid theory predictions 
        are one-parameter fits to the form Eq.~(\protect\ref{eqn:liq}). 
        Curves have been vertically offset for clarity.}
    \label{fig:fig6}
  \end{center}
\end{figure}

        Fig.~\ref{fig:fig6} shows comparisons of the theoretically
predicted scaling of $ \ln (-r^{\frac{1}{2}} \ln g(r)) $ with the values
obtained from our simulations for $\nu = 2$,  $\nu = 1$ and $\nu = 1/2$.
These comparisons involve no adjustable parameters once $n$ has been fixed
(Fig.~\ref{fig:fig5}), since the force scale
$F_o$ was specified for each simulation. 
The results for all three pinning distributions
are in very good agreement with the
form predicted by Eq.~(\ref{eqn:gth1}) for $g(r)$ near lift-off.

The relative success of the present theory compared with the
liquid structure Ansatz in predicting the behavior of $g(r)$
near lift-off emphasizes fundamental differences between the
two classes of systems.
In thermal systems such as simple liquids, structural correlations
arise from the interplay of interaction energies and the
thermal energy scale.
Systems such as pinned flux lines are dominated by quenched disorder
and select configurations in which \emph{forces} are balanced.
Nevertheless, the two approaches yield comparable predictions
for the flux line distribution because of the exponential
form of the pair potential, Eq.~(\ref{eqn:Kasympt}).

        Fig.~\ref{fig:fig4}(a) reveals that a pinscape
generated by identical pins differs from the more broadly
distributed cases in that it gives rise to a distinctively 
steep lift-off in the pair correlation function.
This behavior is captured also by the present treatment. 
In the low-density limit, the theoretical 
description, Eq.~(\ref{eqn:gth1}), 
predicts a step-like increase in $g(r)$ for the smallest 
separation at which
a pair of flux lines can be pinned against their mutual 
repulsion. Fig.~\ref{fig:fig4}(a) shows 
somewhat smoother behavior. Analyzing the 
configurations contributing
to the lift-off of $g(r)$ shows that this smoothing is due to
many-body effects, {\em i.e.} flux lines pinned by the interaction with
two or more close-by flux lines. Comparing this discrepancy to the good
agreement 
with the theoretical prediction in the case of broadly distributed 
pinning strengths leads us to conclude
that, in these cases, the disorder due to variation 
in pinning strength dominates the contribution
arising from the spatial disorder of the flux line 
configurations. 

        A liquid-like form for $g(r)$ can be 
obtained only if the distribution of dynamically 
selected pins is exponential. Eqs.~(\ref{eqn:gth1}) and 
(\ref{eqn:qd}) show 
that, in the general case,
prefactors and different forms of leading 
order behavior can arise depending
on the the tails of the (dynamical) pinning 
energy distribution. Our
results in Fig.~\ref{fig:fig6} suggest, 
however, that this dependence is
weak and will become detectable only 
for very small values of $r$. This is consistent with
the success of Eq.~(\ref{eqn:liq}) in describing
experimental data. We also investigated systems of particles
interacting with power law potentials and found that, when
the interactions fall off sufficiently fast, the pair correlation 
is again of the form Eq.~(\ref{eqn:gth1}).         
        
\section{Discussion and Conclusions}

We have shown that the pair correlation function arising from
stable configurations of interacting particles in a
random pinscape can contain information about both the particle 
interactions and also the statistical properties of the
underlying pinscape.
Our analysis shows that this is the case in the
regime where the pinning and pair interactions are of
comparable strength.
%, with the disorder effectively probing the pair interactions. 
If the pinning completely 
dominates pair interactions, the pair correlation 
function will essentially reveal the spatial disorder of 
the active pinning sites. In the opposite limit, in which 
the pair interactions dominate, the pair correlation 
function will contain features corresponding to crystalline 
order. In both cases, little information can be extracted 
about the form of the interactions.

We have derived an expression for the pair correlation 
function in the limit of point-like disorder and 
low flux line density, Eq.~(\ref{eqn:gth1}). Our 
theory shows that the small $r$ behavior of the pair 
correlation function depends both on the pair 
interaction and also on the availability of strong pinning sites. 
We found that the distribution of the latter is not necessarily 
the same as the distribution of {\em a priori} available 
pinning sites. Rather, the dynamics favor strong pins.  
Inspection of the stable configurations revealed 
that the distribution of dynamically selected pins is consistent 
with a process of choosing the strongest out of $n$ 
available pins. For the parameters used in our 
simulations, $n = 2$ described the dynamically selected 
pinning strength distributions very well.
We expect however that the parameter $n$, while apparently 
independent of the width of the pinning strength distribution, 
should depend on both the concentration of pins and also the penetration
depth of the flux line interaction. A more detailed investigation will be
left for future work.
   
   We now turn to a comparison of the qualitative features of the 
experimentally obtained pair correlation functions with those obtained
from simulation. We find that the width and location of the first peak 
of $g(r)$ is
different: while the first peak of the pair correlation function as 
obtained from  experiments ({\em cf.\ }Fig.~\ref{fig:fig1}) 
is broad and occurs around the mean flux line spacing, 
we see a much narrower peak in our simulations 
({\em cf.\ }Fig.~\ref{fig:fig2}). 
Moreover the 
first peak occurs at a distance significantly 
smaller than the mean flux line
spacing. 
Thus the numerical simulations have larger fluctuations in
the local flux line density than are seen in experiment. A reason for this
is that our particle-like treatment of flux lines ignores their magnetic
properties. One expects that including boundary conditions for
a uniform applied external magnetic field would favor a homogeneous flux line
concentration and thus would penalize the large fluctuations in our
simulations.   

We thank Gene Mazenko for enlightening discussions. This work
was supported in part by the National Science Foundation
through the Science and Technology Center
for Superconductivity under Award Number DMR-9120000 and in part by
the MRSEC program of the National Science Foundation under Award Number
DMR-9400379. C.~H.~S. would like to acknowledge the support
of the National University of Singapore. 
M.~M. acknowledges partial support from the Kerpi\c{s}\c{c}i Foundation. 

\end{document}